\documentclass[letterpaper,12pt,notoc]{JHEP3}
\def\pd{\partial}
\def\mc{\mathcal}

\usepackage{graphicx}
\usepackage{amsmath}
\usepackage{amssymb}

\preprint{ \hbox{}\hfill arXiv: 1407.2762}

\title{N=2 SO(4) 7D gauged supergravity with topological
mass term from 11 dimensions}
\author{Parinya Karndumri\\
String Theory and Supergravity Group, Department
of Physics, Faculty of Science, Chulalongkorn University, 254 Phayathai Road, Pathumwan, Bangkok 10330, Thailand\\
E-mail: \email{parinya.ka@hotmail.com}}

\abstract{We construct a consistent reduction ansatz of
eleven-dimensional supergravity to $N=2$ $SO(4)$ seven-dimensional
gauged supergravity with topological mass term for the three-form
field. The ansatz is obtained from a truncation of the $S^4$
reduction giving rise to the maximal $N=4$ $SO(5)$ gauged
supergravity. Therefore, the consistency is guaranteed by the
consistency of the $S^4$ reduction. Unlike the gauged supergravity
without topological mass having a half-supersymmetric domain wall
vacuum, the resulting 7D gauged supergravity theory admits a
maximally supersymmetric $AdS_7$ critical point. This corresponds to
$N=(1,0)$ superconformal field theory in six dimensions. We also
study RG flows from this $N=(1,0)$ SCFT to non-conformal $N=(1,0)$
Super Yang-Mills theories in the seven-dimensional framework and use
the reduction ansatz to uplift this RG flow to eleven dimensions.}

\begin{document}
\section{Introduction}
Gauged supergravities in various dimensions play an important role
in both string compactifications and in the AdS/CFT correspondence.
In some cases, a consistent truncation can be made in such a way
that a lower dimensional gauged supergravity is obtained via a
dimensional reduction of a (gauged) supergravity in higher
dimensions on spheres \cite{Pope_sphere}. Embedding lower
dimensional gauged supergravities is now of considerable interest
since this provides a method to uplift lower dimensional solutions
to string/M theory.
\\
\indent It is known that sphere reductions of 10 or 11 dimensional
supergravities give rise to gauged supergravity in lower dimensions.
Well-known examples of these consistent sphere reductions include
$S^7$ and $S^4$ reductions of eleven-dimensional supergravity and
$S^5$ reduction of type IIB theory giving rise to $SO(8)$, $SO(5)$
and $SO(6)$ gauged supergravities in four, seven and five
dimensions, respectively
\cite{deWit_S7,S4_reduction_11D,Pope_S5_reduction}. According to the
AdS/CFT correspondence \cite{maldacena}, seven-dimensional gauged
supergravity is useful in the study of $N=(2,0)$ and $N=(1,0)$ field
theories in six dimensions
\cite{LargeN_2_0,Berkooz_6D_dual,AdS7_orbifold1,AdS7_orbifold2,Ferrara_AdS7CFT6}.
The latter describe the dynamics of M5-branes worldvolume in
M-theory and are less-known on the field theory side. Therefore,
seven-dimensional gauged supergravity is expected to give some
insight to six-dimensional field theories via gauge/gravity
correspondence.
\\
\indent In this paper, we are interested in obtaining $N=2$
seven-dimensional gauged supergravity with $SO(4)$ gauged group and
topological mass term. In seven dimensions, the theory is obtained
by coupling three vector multiplets to the pure $SU(2)$ gauged
supergravity constructed in \cite{Pure_N2_7D1}. This matter-coupled
theory has been constructed in \cite{Eric_N2_7D} and \cite{Park_7D}.
The $SO(4)$ gauged supergravity has also been constructed in
\cite{Salam_7DN2} by truncating the maximal $N=4$ $SO(5)$ gauged
supergravity. All of these constructions have not included the
topological mass term for the three-form field, and the resulting
theory does not admit $AdS_7$ vacuum solutions. It has been shown in
\cite{Eric_N2_7Dmassive} that the topological mass term is possible.
The massive gauged theory has been explored in \cite{7D_flow} in
which new $AdS_7$ vacua and the corresponding RG flow interpolating
between these vacua have been given.
\\
\indent To give an interpretation to this solution in the string/M
theory context, it is necessary to embed this solution to 10 or 11
dimensions. The reduction ansatz of eleven-dimensional supergravity
giving rise to pure $SU(2)$ gauged supergravity has been given in
\cite{Pope_N27D_pure}. The $SO(4)$ gauged theory without topological
mass term from a dimensional reduction of eleven- and
ten-dimensional supergravity has been given in
\cite{Salam_Sezgin_from10D} using the result of
\cite{S3_S4_reduction_IIA}. This result is clearly not sufficient to
uplift the solution in \cite{7D_flow}. The dimensionally reduced
theory needs to include the topological mass term in order to admit
$AdS_7$ vacua. We will give an extension to the result of
\cite{Pope_N27D_pure,Salam_Sezgin_from10D} by constructing $SO(4)$
gauged theory including topological mass term from a truncation of
$S^4$ reduction of eleven dimensional supergravity. This provides an
ansatz to uplift the 7-dimensional solutions of massive $N=2$
$SO(4)$ gauged supergravity to eleven dimensions.
\\
\indent The paper is organized as follow. In section
\ref{7D_gaugedN2}, we give relevant formulae for $N=2$ $SO(4)$
gauged supergravity in seven dimensions. The embedding of this
theory in eleven dimensions is obtained via a consistent truncation
of the $S^4$ reduction of eleven-dimensional supergravity in section
\ref{11D_reduciton}. We then use the resulting ansatz to uplift RG
flow solutions from the maximally supersymmetric $AdS_7$ vacuum with
$SO(4)$ symmetry to non-conformal SYM in section
\ref{uplifting_the_solution}. We end the paper by giving some
conclusions and comments in section \ref{conclusion}.

\section{$SO(4)$ $N=2$ gauged supergravity in seven dimensions}\label{7D_gaugedN2}
In this section, we give a description of $SO(4)$ $N=2$ gauged
supergravity in seven dimensions with topological mass term. All of
the notations are the same as those in \cite{Eric_N2_7Dmassive} to
which the reader is referred for further details.
\\
\indent The $SO(4)$ gauged theory is obtained by coupling three
vector multiplets to the $N=2$ supergravity multiplet. The field
contents are given respectively by
\begin{eqnarray}
\textrm{Supergravity multiplet}&:&\qquad (e^a_\mu, \psi^A_\mu,
A^i_\mu,\chi^A,B_{\mu\nu},\sigma) \nonumber \\
\textrm{Vector multiplets}&:&\qquad (A_\mu,\lambda^A,\phi^i)^r
\end{eqnarray}
where an index $r=1,2,3$ labels the three vector multiplets. Curved
and flat space-time indices are denoted by $\mu,\nu,\ldots$ and
$a,b,\ldots$, respectively. $B_{\mu\nu}$ and $\sigma$ are a two-form
and the dilaton fields. The two-form field will be dualized to a
three-form field $C_{\mu\nu\rho}$. Indices $i,j=1,2,3$ label
triplets of $SU(2)_R$. The $9$ scalars $\phi^{ir}$ are parametrized
by $SO(3,3)/SO(3)\times SO(3)\sim SL(4,\mathbb{R})/SO(4)$ coset
manifold. The corresponding coset representative of
$SO(3,3)/SO(3)\times SO(3)$ will be denoted by
\begin{equation}
L=(L_I^{\phantom{s}i},L_I^{\phantom{s}r}), \qquad I=1,\ldots, 6\, .
\end{equation}
whose inverse is given by
$L^{-1}=(L^I_{\phantom{s}i},L^I_{\phantom{s}r})$ where
$L^I_{\phantom{s}i}=\eta^{IJ}L_{Ji}$ and
$L^I_{\phantom{s}r}=\eta^{IJ}L_{Jr}$. Indices $i,j$ and $r,s$ are
raised and lowered by $\delta_{ij}$ and $\delta_{rs}$, respectively
while the full $SO(3,3)$ indices $I,J$ are raised and lowered by
$\eta_{IJ}=\textrm{diag}(---+++)$.
\\
\indent The $SO(4)\sim SU(2)\times SU(2)$ gauging is implemented by
promoting the $SU(2)\times SU(2) \sim SO(3)\times SO(3)\subset
SO(3,3)$ to a gauge symmetry. The structure constants for the
$SU(2)\times SU(2)$ gauge group, which will appear in various
quantities, are given by
\begin{equation}
f_{IJK}=(g_1\epsilon_{ijk},g_2 \epsilon_{rst}).
\end{equation}
To obtain $SO(4)$ gauge group, we will later set $g_2=g_1$. The
bosonic Lagrangian can be written in a form language as
\begin{eqnarray}
\mc{L}&=&\frac{1}{2}R*\mathbb{I}-\frac{1}{2}e^\sigma a_{IJ}*F^I_{(2)}\wedge F^J_{(2)}-\frac{1}{2}e^{-2\sigma}*H_{(4)}\wedge H_{(4)}-\frac{5}{8}*d \sigma \wedge d\sigma\nonumber \\
& &-\frac{1}{2}*P^{ ir}\wedge P_{ir}+\frac{1}{\sqrt{2}}H_{(4)}\wedge
\omega_{(3)}-4hH_{(4)}\wedge C_{(3)}-V*\mathbb{I}\label{7Daction}
\end{eqnarray}
where the scalar potential is given by
\begin{equation}
V=\frac{1}{4}e^{-\sigma}\left(C^{ir}C_{ir}-\frac{1}{9}C^2\right)+16h^2e^{4\sigma}
-\frac{4\sqrt{2}}{3}he^{\frac{3\sigma}{2}}C\,.
\end{equation}
The constant $h$ describes the topological mass term for the
three-form $C_{(3)}$ with $H_{(4)}=dC_{(3)}$. The quantities
appearing in the above Lagrangian are defined by
\begin{eqnarray}
P_\mu^{ir}&=&L^{Ir}\left(\delta^K_I\pd_\mu+f_{IJ}^{\phantom{sad}K}A_\mu^J\right)L^i_{\phantom{s}K},
\qquad
C_{rsi}=f_{IJ}^{\phantom{sad}K}L^I_{\phantom{s}r}L^J_{\phantom{s}s}L_{Ki},\nonumber
\\
C_{ir}&=&\frac{1}{\sqrt{2}}f_{IJ}^{\phantom{sad}K}L^I_{\phantom{s}j}L^J_{\phantom{s}k}L_{Kr}\epsilon^{ijk},
\qquad
C=-\frac{1}{\sqrt{2}}f_{IJ}^{\phantom{sad}K}L^I_{\phantom{s}i}L^J_{\phantom{s}j}L_{Kk}\epsilon^{ijk},\nonumber \\
a_{IJ}&=&L^i_{\phantom{s}I}L_{iJ}+L^r_{\phantom{s}I}L_{rJ}\, .
\end{eqnarray}
The Chern-Simons three-form satisfying
$d\omega_{(3)}=F^I_{(2)}\wedge F^I_{(2)}$ is given by
\begin{equation}
\omega_{(3)}=F^I_{(2)}\wedge
A^I_{(1)}-\frac{1}{6}f_{IJ}^{\phantom{sa}K}A^I_{(1)}\wedge
A^J_{(1)}\wedge A_{(1)K}
\end{equation}
with
$F^I_{(2)}=dA^I_{(1)}+\frac{1}{2}f_{JK}^{\phantom{sas}I}A^J_{(1)}\wedge
A^K_{(1)}$
\\
\indent It is also useful to give the corresponding field equations
\begin{eqnarray}
d\left(e^{-2\sigma}*H_{(4)}\right)+8h H_{(4)}-\frac{1}{\sqrt{2}}F^I_{(2)}\wedge F^I_{(2)}&=&0,\label{7D4form_eq}\\
\frac{5}{4}d*d\sigma-\frac{1}{2}e^{\sigma}a_{IJ}*F^I_{(2)}\wedge F^J_{(2)}+e^{-2\sigma}*H_{(4)}\wedge H_{(4)}\qquad & &\nonumber \\
+\left[\frac{1}{4}e^{-\sigma}\left(C^{ir}C_{ir}-\frac{1}{2}C^2\right)+2\sqrt{2}he^{\frac{3}{2}\sigma}C-64 h^2 e^{4\sigma}\right]\epsilon_{(7)}&=&0\label{7Ddilato_eq}\\
D(e^\sigma a_{IJ}*F^I_{(2)})-\sqrt{2}H_{(4)}\wedge F^J_{(2)}+*P^{ir}f_{IJ}^{\phantom{sas}K}L^I_{\phantom{s}r}L_{iK}&=&0\label{7DYM_eq}\\
D*P^{ir}-2e^\sigma
L^i_{\phantom{s}I}L^r_{\phantom{s}J}*F^I_{(2)}\wedge
F^J_{(2)}\qquad\qquad
& &\nonumber \\
-*\mathbb{I}
\left[\frac{1}{\sqrt{2}}e^{-\sigma}C_{jr}C^{rsk}\epsilon^{ijk}+4\sqrt{2}he^{\frac{3\sigma}{2}}C_{ir}\right]&=&0\,
.\label{7Dscalar_eq}
\end{eqnarray}
The Yang-Mills equation \eqref{7DYM_eq} can be written in terms of
$C^{ir}$ and $C^{irs}$ by using the relation
\begin{equation}
f_{IJ}^{\phantom{sad}K}L^I_{\phantom{s}r}L_{iK}=-\frac{1}{2\sqrt{2}}\epsilon^{ijk}C^{jr}L^k_{\phantom{s}J}-
C^{irs}L_{sJ}\, .
\end{equation}
In obtaining the scalar equation \eqref{7Dscalar_eq}, we have used
the projections in the variations of scalars as in \cite{Eric_N2_7D}
\begin{eqnarray}
\delta L^i_{\phantom{s}I}&=&X^i_{\phantom{s}r}L^r_{\phantom{s}I}+X^i_{\phantom{s}j}L^j_{\phantom{s}I},\nonumber \\
\delta L^r_{\phantom{s}I}&=&{X^r}_s{L^s}_I+{X^r}_i{L^i}_I
\end{eqnarray}
which lead to
\begin{eqnarray}
\delta C^2&=&-6\sqrt{2}CC^{ir}X_{ir},\nonumber \\
\delta
(C^{ir}C_{ir})&=&2\sqrt{2}C_{js}C^{rsk}\epsilon^{ijk}{X^i}_r-\frac{2\sqrt{2}}{3}C_{ir}CX^i_{\phantom{s}r}\,
.
\end{eqnarray}
\indent We finally give supersymmetry transformations for fermions
with all fermionic fields vanishing. These are given by
\begin{eqnarray}
\delta \psi_\mu &=&2D_\mu
\epsilon-\frac{\sqrt{2}}{30}e^{-\frac{\sigma}{2}}C\gamma_\mu
\epsilon
-\frac{1}{240\sqrt{2}}e^{-\sigma}H_{\rho\sigma\lambda\tau}\left(\gamma_\mu
\gamma^{\rho\sigma\lambda
\tau}+5\gamma^{\rho\sigma\lambda\tau}\gamma_\mu\right)\epsilon\nonumber
\\
&
&-\frac{i}{20}e^{\frac{\sigma}{2}}F^i_{\rho\sigma}\sigma^i\left(3\gamma_\mu
\gamma^{\rho\sigma}-5\gamma^{\rho\sigma}\gamma_\mu\right)\epsilon
-\frac{4}{5}he^{2\sigma}\gamma_\mu \epsilon,\label{delta_psi}\\
\delta \chi &=&-\frac{1}{2}\gamma^\mu\pd_\mu \sigma
\epsilon-\frac{i}{10}e^{\frac{\sigma}{2}}F^i_{\mu\nu}\sigma^i\gamma^{\mu\nu}\epsilon-
\frac{1}{60\sqrt{2}}e^{-\sigma}H_{\mu\nu\rho\sigma}\gamma^{\mu\nu\rho\sigma}\epsilon\nonumber
\\
& &+\frac{\sqrt{2}}{30}e^{-\frac{\sigma}{2}}C\epsilon-\frac{16}{5}e^{2\sigma}h\epsilon,\label{delta_chi}\\
\delta \lambda^r &=&-i\gamma^\mu
P^{ir}_\mu\sigma^i\epsilon-\frac{1}{2}e^{\frac{\sigma}{2}}F^r_{\mu\nu}\gamma^{\mu\nu}\epsilon-\frac{i}{\sqrt{2}}e^{-\frac{\sigma}{2}}C^{ir}\sigma^i\epsilon\label{delta_lambda}
\end{eqnarray}
where $SU(2)_R$ doublet indices $A,B,\ldots$ on spinors are
suppressed. $\sigma^i$ are the usual Pauli matrices.

\section{Seven dimensional $N=2$ gauged supergravity from eleven dimensions}\label{11D_reduciton}
We now construct a reduction ansatz for embedding $SO(4)$ $N=2$
gauged supergravity mentioned in the previous section in eleven
dimensions. The ansatz will be obtained from a consistent truncation
of the $S^4$ reduction of eleven-dimensional supergravity giving
rise to the maximal $N=4$ $SO(5)$ gauged supergravity in seven
dimensions. To obtain the topological mass term, we will impose the
so-called odd-dimensional self-duality as in \cite{Pope_N27D_pure}.

\subsection{$N=4$ $SO(5)$ gauged supergravity from seven dimensions}
To set up the notations and make the paper self-contained, we
briefly repeat the $S^4$ reduction of eleven-dimensional
supergravity \cite{S4_reduction_11D,Peter_S4}. We will work in the
notations of \cite{S3_S4_reduction_IIA} and deal mainly with bosonic
fields. The field content of eleven-dimensional supergravity
consists of the graviton $\hat{g}_{MN}$, gravitino $\hat{\psi}_M$
and a four-form field $\hat{F}_{(4)}$. Eleven-dimensional space-time
indices are denoted by $M,N=0,1,\ldots, 10$.
\\
\indent The $S^4$ reduction is characterized by the following ansatz
\begin{eqnarray}
d\hat{s}^2_{11}&=&\Delta^{\frac{1}{3}}ds^2_7+\frac{1}{g^2}\Delta^{-\frac{2}{3}}T^{-1}_{ij}D\mu^i D\mu^j,\\
\hat{F}_{(4)}&=&\frac{1}{4!}\epsilon_{i_1\ldots i_5}\left[\frac{4}{g^3}\Delta^{-2}\mu^m\mu^nT^{i_1m}DT^{i_2n}\wedge D\mu^{i_3}\wedge D\mu^{i_4}\wedge D\mu^{i_5}\right. \nonumber \\
& &\left.+\frac{6}{g^2}\Delta^{-1}T^{i_5j}\mu^jF_{(2)}^{i_1i_2}\wedge D\mu^{i_3}\wedge D\mu^{i_4}-\frac{1}{g^3}\Delta^{-2}U\mu^{i_1}D\mu^{i_2}\wedge \ldots \wedge D\mu^{i_5}\right]\nonumber \\
& &-T_{ij}*S^i_{(3)}\mu^j+\frac{1}{g}S^i_{(3)}\wedge
D\mu^i
\end{eqnarray}
where the quantities appearing in the above equations are defined by
\begin{eqnarray}
U&=&2T_{ij}T_{jk}\mu^i\mu^k-\Delta T_{ii},\qquad \Delta=T_{ij}\mu^i\mu^j,\qquad \mu^i\mu^i=1,\nonumber \\
F^{ij}_{(2)}&=&dA^{ij}_{(1)}+gA^{ik}_{(1)}\wedge A^{kj}_{(1)},\qquad
D\mu^i=d\mu^i+gA^{ij}_{(1)}\mu^j,\nonumber \\
DT_{ij}&=&dT_{ij}+gA^{ik}_{(1)}T_{kj}+gA^{jk}_{(1)}T_{ik}\, .
\end{eqnarray}
The symmetric matrix $T_{ij}$, $i,j=1,\ldots ,5$ with unit
determinant parametrize the $SL(5,\mathbb{R})/SO(5)$ coset manifold.
\\
\indent The bosonic field content of $N=4$ gauged supergravity is
given by the metric $g_{\mu\nu}$, ten vectors
$A^{ij}_{(1)}=A^{[ij]}_{(1)}$ gauging the $SO(5)$ gauge group, five
three-form fields $S^i_{(3)}$ and four-teen scalars $T_{ij}$. The
corresponding field equations are given by
\begin{eqnarray}
D(T_{ij}*S^j_{(3)})&=&F^{ij}_{(2)}\wedge S^j_{(3)},\label{3-form_eq1}\\
H^i_{(4)}&=&gT_{ij}*S^j_{(3)}+\frac{1}{8}\epsilon_{ij_1\ldots j_4}F^{j_1j_2}_{(2)}\wedge F^{j_3j_4}_{(2)},\label{3-form_eq2}\\
D(T^{-1}_{ik}T^{-1}_{jl}*F^{ij}_{(2)})&=&-2gT^{-1}_{i[k}*DT_{l]i}-\frac{1}{2g}\epsilon_{i_1i_2i_3kl}F^{i_1i_2}_{(2)}\wedge H^{i_3}_{(4)}\nonumber \\
& &+\frac{3}{2g}\delta^{j_1j_2j_3j_4}_{i_1i_2kl}F^{i_1i_2}_{(2)}\wedge F^{j_1j_2}_{(2)}\wedge F^{j_3j_4}_{(2)}-S^k_{(3)}\wedge S^l_{(3)},\label{YM_eq}\\
D(T^{-1}_{ik}*DT_{kj})&=&2g^2\left(2T_{ik}T_{kj}-T_{kk}T_{ij}\right)\epsilon_{(7)}+T_{im}^{-1}T^{-1}_{kl}
*F^{ml}_{(2)}\wedge F^{kj}_{(2)}\nonumber \\
& &+T_{jk}*S^k_{(3)}\wedge S^i_{(3)}-\frac{1}{5}\delta_{ij}\left[2g^2\left(2T_{kl}T_{kl}-(T_{kk})^2\right)\epsilon_{(7)}\right. \nonumber \\
& &\left.+T^{-1}_{nm}T^{-1}_{kl}*F^{ml}_{(2)}\wedge
F^{kn}_{(2)}+T_{kl}*S^k_{(3)}\wedge
S^l_{(3)}\right]\label{scalar_eq}
\end{eqnarray}
where
\begin{equation}
H^i_{(4)}=DS^i_{(3)}=dS^i_{(3)}+gA^{ij}_{(1)}\wedge S^j_{(3)}\, .
\end{equation}
All of these equation can be obtained from the Lagrangian
\begin{eqnarray}
\mc{L}_7&=&R*\mathbb{I}-\frac{1}{4}T^{-1}_{ij}*DT_{jk}\wedge
T^{-1}_{kl}DT_{li}
-\frac{1}{4}T^{-1}_{ik}T^{-1}_{jl}*F^{ij}_{(2)}\wedge F^{kl}_{(2)}
-\frac{1}{4}T_{ij}*S^i_{(3)}\wedge S^j_{(3)}\nonumber \\
& &+\frac{1}{2g}S^i_{(3)}\wedge
H^i_{(4)}-\frac{1}{8g}\epsilon_{i_{j_1\ldots j_4}}S^i_{(3)}\wedge
F^{j_1j_2}_{(2)}\wedge
F^{j_3j_4}_{(2)}+\frac{1}{g}\Omega_{(7)}-V*\mathbb{I}
\end{eqnarray}
where $\Omega_{(7)}$ is the Chern-Simens three-form whose explicit
form can be found in \cite{7DN4_Peter}. The scalar potential for
$T_{ij}$ is given by
\begin{equation}
V=g^2\left(T_{ij}T_{ij}-\frac{1}{2}(T_{ii})^2\right).
\end{equation}
\indent We have not given Einstein equation since we will not
consider Einstein equation in this paper. The consistency of the
full truncation, including the Einstein equation, to $N=2$ $SO(4)$
gauged supergravity is guaranteed from the consistency of the $S^4$
reduction.
\\
\indent For completeness, we also repeat supersymmetry
transformations of fermionic fields $\psi_\mu$ and
$\lambda_{\hat{i}}$. Indices $\hat{i},\hat{j}=1,\ldots, 5$ are
vector indices of the composite $SO(5)_c$ symmetry. Additionally,
both $\psi_\mu$ and $\lambda_{\hat{i}}$ transform as a spinor under
$SO(5)_c$ with the condition $\Gamma^{\hat{i}}\lambda_{\hat{i}}=0$,
but we have omitted the $SO(5)_c$ spinor indices to make the
following expressions more compact. The $SO(5)_c$ gamma matrices
will be denoted by $\Gamma^{\hat{i}}$. The associated supersymmetry
transformations are given by \cite{7DN4_Peter}
\begin{eqnarray}
\delta \psi_\mu &=&D_\mu \epsilon
-\frac{1}{20}gT_{\hat{i}\hat{i}}\gamma_\mu
\epsilon-\frac{1}{40\sqrt{2}}\left(\gamma_\mu^{\phantom{s}\nu\rho}-8\delta^\nu_\mu\gamma^\rho\right)
F^{\hat{i}\hat{j}}_{\nu\rho}\Gamma_{\hat{i}\hat{j}}\epsilon \nonumber \\
&
&-\frac{1}{60}\left(\gamma_\mu^{\phantom{s}\nu\rho\sigma}-\frac{9}{2}\delta^\nu_\mu
\gamma^{\rho\sigma}\right)S_{\hat{i} \nu\rho\sigma}
\Gamma^{\hat{i}}\epsilon ,\label{deltaPsiN4}\\
\delta
\lambda_{\hat{i}}&=&\frac{1}{16\sqrt{2}}\gamma^{\mu\nu}\left(\Gamma_{\hat{k}\hat{l}}\Gamma_{\hat{i}}
-\frac{1}{5}\Gamma_{\hat{i}}\Gamma_{\hat{k}\hat{l}}\right)F^{\hat{k}\hat{l}}_{\mu\nu}\epsilon +\frac{1}{2}\gamma^\mu \Gamma^{\hat{j}}P_{\mu\hat{i}\hat{j}}\epsilon \nonumber \\
&
&-\frac{1}{120}\gamma^{\mu\nu\rho}\left(\Gamma_{\hat{i}}^{\phantom{s}\hat{j}}-4\delta^{\hat{j}}_{\hat{i}}\right)
S_{\hat{j}\mu\nu\rho}\epsilon
+\frac{1}{2}g\left(T_{\hat{i}\hat{j}}-\frac{1}{5}T_{\hat{k}\hat{k}}\delta_{\hat{i}\hat{j}}
\right)\Gamma^{\hat{j}}\epsilon\label{deltaLambdaN4}
\end{eqnarray}
where
\begin{eqnarray}
F^{\hat{i}\hat{j}}_{(2)}&=&\Pi_{i}^{\phantom{s}\hat{i}}\Pi_{j}^{\phantom{s}\hat{j}}F^{ij}_{(2)},\qquad
T_{\hat{i}\hat{j}}=(\Pi^{-1})_{\hat{i}}^{\phantom{s}i}(\Pi^{-1})_{\hat{j}}^{\phantom{s}j}\delta^{ij},\nonumber \\
D\epsilon &=& d\epsilon +\frac{1}{4}\omega_{ab}\gamma^{ab}\epsilon +\frac{1}{4}Q_{\hat{i}\hat{j}}
\Gamma^{\hat{i}\hat{j}}\epsilon,\qquad
T^{ij}=(\Pi^{-1})_{\hat{i}}^{\phantom{s}i}(\Pi^{-1})_{\hat{j}}^{\phantom{s}j}\delta^{\hat{i}\hat{j}},\nonumber \\
P_{(\hat{i}\hat{j})}+Q_{[\hat{i}\hat{j}]}&=&(\Pi^{-1})
_{\hat{i}}^{\phantom{s}i}\left(\delta^j_id
+gA_{(1)i}^{\phantom{ssas}j}\right)\Pi
_{j}^{\phantom{s}\hat{k}}\delta_{\hat{j}\hat{k}},\qquad
S_{(3)\hat{i}}=(\Pi^{-1})_{\hat{i}}^{\phantom{s}i}S_{(3)i}
\end{eqnarray}
with $\Pi _{i}^{\phantom{s}\hat{i}}$ being the
$SL(5,\mathbb{R})/SO(5)$ coset representative.

\subsection{$SO(4)$ $N=2$ gauged supergravity from $S^4$ reduction}
We now truncate the $N=4$ gauged supergravity to $N=2$ theory with
topological mass term for the three-form field and $SO(4)$ gauge
group. In this process, the gauge group $SO(5)$ is broken to
$SO(4)$. We will split the index $i$ as $(\alpha,5)$ with
$\alpha=1,\ldots, 4$. Furthermore, we will set $T_{5\alpha}$,
$S^\alpha$ and $F^{5\alpha}$ to zero. The $S^4$ coordinates $\mu^i$
will be chosen to be $\mu^i=(\cos \xi \mu^\alpha,\sin \xi)$ in which
$\mu^\alpha$ satisfy $\mu^\alpha \mu^\alpha=1$. Similar to $\mu^i$,
$\mu^\alpha$ are coordinates on $S^3$. The scalar truncation is
given by
$T_{ij}=(T_{\alpha\beta},T_{55})=(X\tilde{T}_{\alpha\beta},X^{-4})$
with $\tilde{T}_{\alpha\beta}$ being unimodular. The scalar field
$X$ will be related to the $N=2$ dilaton.
\\
\indent With these truncations, the three-form field equations
\eqref{3-form_eq1} and \eqref{3-form_eq2} become
\begin{eqnarray}
D(X^{-4}*S^5_{(3)})&=&0\label{3-form_eq11}\\
dS^5_{(3)}&=&gX^{-4}*S^5_{(3)}+\frac{1}{8}\epsilon_{\alpha\beta\gamma\delta}F^{\alpha\beta}_{(2)}\wedge
F^{\gamma\delta}_{(2)}\, .\label{3-form_eq22}
\end{eqnarray}
We have used
$\epsilon_{5\alpha\beta\gamma\delta}=\epsilon_{\alpha\beta\gamma\delta}$.
From \eqref{3-form_eq11}, we see that the four-form
$X^{-4}*S^5_{(3)}$ is closed. We will denote it by
\begin{equation}
X^{-4}*S^5_{(3)}=-F_{(4)}=-dC_{(3)}
\end{equation}
or
\begin{equation}
S^5_{(3)}=X^4*F_{(4)}\, .
\end{equation}
To satisfy equation \eqref{3-form_eq22}, we impose the
odd-dimensional self-duality condition
\begin{equation}
S^5_{(3)}=-gC_{(3)}+\omega_{(3)}
\end{equation}
or
\begin{equation}
X^4*F_{(4)}=-gC_{(3)}+\omega_{(3)}
\end{equation}
where $\omega_{(3)}$, satisfying
$d\omega_{(3)}=\frac{1}{8}\epsilon_{\alpha\beta\gamma\delta}F^{\alpha\beta}_{(2)}\wedge
F^{\gamma\delta}_{(2)}$, is the Chern-Simons term given by
\begin{equation}
\omega_{(3)}=\frac{1}{8}\epsilon_{\alpha\beta\gamma\delta}\left(F^{\alpha\beta}_{(2)}\wedge
A^{\gamma\delta}_{(1)}-\frac{1}{3}gA^{\alpha\beta}_{(1)}\wedge
A^{\gamma\kappa}_{(1)}\wedge A^{\kappa\delta}_{(1)}\right).
\end{equation}
Equations for $S^\alpha_{(3)}$ are trivially satisfied.
\\
\indent For the Yang-Mills equations, it can be verified that
setting $F^{5\alpha}_{(2)}=0$ satisfies their field equations. For
$F^{\alpha\beta}_{(2)}$, we find
\begin{equation}
D\left(X^{-2}\tilde{T}^{-1}_{\alpha\gamma}\tilde{T}^{-1}_{\beta\delta}*F^{\gamma\delta}_{(2)}\right)=
-2g\tilde{T}^{-1}_{{\gamma[\alpha}}*D\tilde{T}_{\beta]\gamma}+\frac{1}{2}\epsilon_{\alpha\beta\gamma\delta}
F^{\gamma\delta}_{(2)}\wedge F_{(4)}
\end{equation}
where we have used the odd-dimensional self-duality condition.
\\
\indent We then consider scalar equations. Equations for
$T_{5\alpha}$ are trivially satisfied while the $T_{55}$ equation
gives rise to the dilaton eqiation
\begin{eqnarray}
d(X^{-1}*dX)&=&\frac{1}{5}X^4*F_{(4)}\wedge F_{(4)}-\frac{1}{20}X^{-2}\tilde{T}^{-1}_{\alpha\beta}\tilde{T}^{-1}_{\gamma\delta}*F^{\beta\delta}_{(2)}\wedge F^{\alpha\gamma}_{(2)}\nonumber \\
&
&-\frac{1}{10}g^2\left[4X^{-8}-3X^{-3}\tilde{T}_{\alpha\alpha}-2X^2
\left(\tilde{T}_{\alpha\beta}\tilde{T}_{\alpha\beta}-\frac{1}{2}(\tilde{T}_{\alpha\alpha})^2\right)\right]
\epsilon_{(7)}.
\qquad \label{X_eq}
\end{eqnarray}
For $T_{ij}=T_{\alpha\beta}$, we find
\begin{eqnarray}
D(\tilde{T}^{-1}_{\alpha\gamma}*D\tilde{T}_{\gamma\beta})&+&\delta_{\alpha\beta}d(X^{-1}*dX)=
X^{-2}\tilde{T}^{-1}_{\alpha\gamma}\tilde{T}^{-1}_{\delta\kappa}*F^{\gamma\kappa}_{(2)}\wedge F^{\delta\beta}_{(2)} \nonumber \\
&
&+2g^2\left[X^2\left(2\tilde{T}_{\alpha\gamma}\tilde{T}_{\gamma\beta}
-\tilde{T}_{\gamma\gamma}\tilde{T}_{\alpha\beta}\right)-X^{-3}\tilde{T}_{\alpha\beta}\right]\epsilon_{(7)}
\nonumber \\
& &+\delta_{\alpha\beta}\left[\frac{1}{5}X^4*F_{(4)}\wedge
F_{(4)}-\frac{1}{5}X^{-2}\tilde{T}^{-1}_{\gamma\delta}\tilde{T}^{-1}_{\kappa\lambda}
*F^{\delta\lambda}_{(2)}\wedge F^{\kappa\gamma}_{(2)}\right.\nonumber \\
&
&\left.-\frac{2}{5}g^2\left[2X^2\left(\tilde{T}_{\gamma\delta}\tilde{T}_{\gamma\delta}
-\frac{1}{2}(\tilde{T}_{\gamma\gamma})^2\right)+X^{-8}-2X^{-3}\tilde{T}_{\gamma\gamma}\right]\epsilon_{(7)}\right].
\qquad
\end{eqnarray}
We can now use the $X$ equation \eqref{X_eq} and end up with
\begin{eqnarray}
D(\tilde{T}^{-1}_{\alpha\gamma}*D\tilde{T}_{\gamma\beta})&=&2g^2\left[2X^2\left(\tilde{T}_{\alpha\gamma}\tilde{T}_{\gamma\beta}
-\frac{1}{2}\tilde{T}_{\gamma\gamma}\tilde{T}_{\alpha\beta}\right)-X^{-3}\tilde{T}_{\alpha\beta}\right]\epsilon_{(7)}
\nonumber \\
& &+
X^{-2}\tilde{T}^{-1}_{\alpha\gamma}\tilde{T}^{-1}_{\delta\kappa}*F^{\gamma\kappa}_{(2)}\wedge
F^{\delta\beta}_{(2)}+\delta_{\alpha\beta}\left[\left\{\frac{5}{2}g^2X^2\left(\tilde{T}_{\gamma\delta}\tilde{T}_{\gamma\delta}
-\frac{1}{2}(\tilde{T}_{\gamma\gamma})^2\right)\right.\right.\nonumber \\
& &\left.\left.+\frac{1}{2}g^2 X^{-3}\tilde{T}_{\gamma\gamma}\right\}\epsilon_{(7)}
-\frac{1}{4}X^{-2}\tilde{T}^{-1}_{\gamma\delta}\tilde{T}^{-1}_{\kappa\lambda}*F^{\delta\lambda}_{(2)}\wedge
F^{\kappa\gamma}_{(2)}\right]\label{Tt_eq}
\end{eqnarray}
\indent With all of the above truncations, we find the following
ansatz for the metric and the four-form field
\begin{eqnarray}
d\hat{s}^2_{11}&=&\Delta^{\frac{1}{3}}ds^2_7+\frac{2}{g^2}\Delta^{-\frac{2}{3}}X^3\left[X\cos^2\xi
+X^{-4}\sin^2\xi \tilde{T}^{-1}_{\alpha\beta}\mu^\alpha\mu^\beta\right]d\xi^2\nonumber \\
& &-\frac{1}{g^2}\Delta^{-\frac{2}{3}}X^{-1}\tilde{T}^{-1}_{\alpha\beta}\sin \xi \mu^\alpha d\xi D\mu^\beta+\frac{1}{2g^2}\Delta^{-\frac{2}{3}}X^{-1}\tilde{T}^{-1}_{\alpha\beta}\cos^2\xi D\mu^\alpha D\mu^\beta,\\
\hat{F}_{(4)}&=&F_{(4)}\sin\xi+\frac{1}{g}X^4\cos \xi *F_{(4)}\wedge d\xi +\frac{1}{g^3}\Delta^{-2}U\cos^5\xi d\xi \wedge \epsilon_{(3)}\nonumber \\
&
&+\frac{1}{3!g^3}\epsilon_{\alpha\beta\gamma\delta}\Delta^{-2}X^{-3}\sin\xi
\cos^4\xi \mu^\kappa
\left[5\tilde{T}^{\alpha\kappa}X^{-1}dX+D\tilde{T}^{\alpha\kappa}\right]\wedge
D\mu^\beta\wedge D\mu^\gamma
\wedge D\mu^\delta\nonumber \\
&
&+\frac{1}{2g^3}\epsilon_{\alpha\beta\gamma\delta}\Delta^{-2}\cos^3\xi
\mu^\kappa\mu^\lambda \left[
\cos^2\xi X^2 \tilde{T}^{\alpha\kappa}D\tilde{T}^{\beta\lambda}-\sin^2\xi X^{-3}\delta^{\beta\lambda}D\tilde{T}^{\alpha\kappa}\right.\nonumber \\
& &\left.-5\sin^2\xi \tilde{T}^{\alpha\kappa}X^{-4}\delta^{\beta\lambda}dX\right]\wedge D\mu ^\gamma\wedge D\mu^\delta \wedge d\xi+\frac{1}{2g^2}\cos\xi\epsilon_{\alpha\beta\gamma\delta}\times \nonumber \\
& &\left[\frac{1}{2}\cos\xi \sin \xi X^{-4}D\mu^\gamma-\left(X^{-4}\sin^2\xi \mu^\gamma+X^2\cos^2 \xi \tilde{T}^{\gamma\kappa}\mu^\kappa\right)d\xi\right]\wedge F^{\alpha\beta}_{(2)}\wedge D\mu^\delta\nonumber \\
\end{eqnarray}
where
\begin{eqnarray}
U&=&\sin^2\xi
\left(X^{-8}-X^{-3}\tilde{T}_{\alpha\alpha}\right)+\cos^2\xi
\mu^\alpha\mu^\beta\left(
2X^2\tilde{T}_{\alpha\gamma}\tilde{T}_{\gamma\beta}-X^2\tilde{T}_{\alpha\beta}\tilde{T}_{\gamma\gamma}
-X^{-3}\tilde{T}_{\alpha\beta}\right)\nonumber \\
\epsilon_{(3)}&=&\frac{1}{3!}\epsilon_{\alpha\beta\gamma\delta}\mu^\alpha
D\mu^\beta \wedge D\mu^\gamma \wedge D\mu^\delta\, .
\end{eqnarray}
\\
\indent All of the above equations reduce to the pure $N=2$ gauged
supergravity with $SU(2)$ gauge group for
$\tilde{T}_{\alpha\beta}=\delta_{\alpha\beta}$ after using various
relations given in \cite{Natase_S4}. Note that for
$\tilde{T}_{\alpha\beta}=\delta_{\alpha\beta}$, equation
\eqref{Tt_eq} gives
\begin{equation}
*F^{\alpha\gamma}_{(2)}\wedge F^{\gamma\beta}_{(2)}=\frac{1}{4}\delta_{\alpha\beta}*F^{\gamma\delta}_{(2)}\wedge F^{\delta\gamma}_{(2)}
\end{equation}
which means that the $SO(4)$ gauge fields $A^{\alpha\beta}_{(1)}$
must be truncated to those of $SU(2)$ satisfying
$F^{\alpha\beta}_{(2)}=\pm\frac{1}{2}\epsilon_{\alpha\beta\gamma\delta}F^{\gamma\delta}_{(2)}$.
This is expected since there are only three vector fields in the
pure gauged supergravity which only admit $SU(2)$ gauging.
\\
\indent The above equations can be obtained from the Lagrangian
\begin{eqnarray}
\mc{L}_7&=&R*\mathbb{I}-\frac{1}{4}X^{-2}\tilde{T}^{-1}_{\alpha\gamma}\tilde{T}^{-1}_{\beta\delta}
*F^{\alpha\beta}_{(2)}
\wedge
F^{\gamma\delta}_{(2)}-\frac{1}{4}\tilde{T}^{-1}_{\alpha\beta}*D\tilde{T}_{\beta\gamma}\wedge
\tilde{T}^{-1}_{\gamma\delta}D\tilde{T}_{\delta\alpha}\nonumber \\
& &-\frac{1}{2}X^4*F_{(4)}\wedge F_{(4)}+\frac{1}{8}\epsilon_{\alpha\beta\gamma\delta}C_{(3)}\wedge F^{\alpha\beta}_{(2)}\wedge F^{\gamma\delta}_{(2)}-5X^{-2}*dX\wedge dX\nonumber \\
& &-\frac{1}{2}gF_{(4)}\wedge C_{(3)}-V*\mathbb{I}\label{L_7D_from_11D}
\end{eqnarray}
where the scalar potential is given by
\begin{equation}
V=\frac{1}{2}g^2\left[X^{-8}-2X^{-3}\tilde{T}_{\alpha\alpha}+2X^2\left(\tilde{T}_{\alpha\beta}
\tilde{T}_{\alpha\beta}-\frac{1}{2}\tilde{T}^2_{\alpha\alpha}\right)\right].\label{potential_11D}
\end{equation}
For $\tilde{T}_{\alpha\beta}=\delta_{\alpha\beta}$, we find $\tilde{T}_{\alpha\alpha}=\tilde{T}_{\alpha\beta}\tilde{T}_{\alpha\beta}=4$. The above potential becomes
\begin{equation}
V=\frac{1}{2}g^2\left(X^{-8}-8X^{-3}-8X^2\right)
\end{equation}
which is exactly the same as that given in \cite{Pope_N27D_pure} up to a redefinition of the coupling constant $g$.
\\
\indent We can also check another truncation namely to $U(1)\times U(1)$ gauged supergravity. To
preserve $SO(2)\times SO(2)$ symmetry, we take the scalar matrix to be
\begin{equation}
\tilde{T}_{\alpha\beta}=\left(
                     \begin{array}{cccc}
                       e^{\frac{\phi_1}{\sqrt{2}}} &  &  &  \\
                        & e^{\frac{\phi_1}{\sqrt{2}}} &  &  \\
                        &  & e^{-\frac{\phi_1}{\sqrt{2}}} &  \\
                        &  &  & e^{-\frac{\phi_1}{\sqrt{2}}} \\
                     \end{array}
                   \right)
\end{equation}
and define $X=e^{-\frac{\phi_2}{\sqrt{10}}}$. The potential \eqref{potential_11D} becomes
\begin{equation}
V=\frac{1}{2}g^2\left[e^{\frac{8\phi_2}{\sqrt{10}}}-8e^{-\frac{2\phi_2}{\sqrt{10}}}-4e^{\frac{3\phi_2}{\sqrt{10}}}
\left(e^{\frac{\phi_1}{\sqrt{2}}}+e^{-\frac{\phi_1}{\sqrt{2}}}\right)\right]
\end{equation}
which takes the same form as that given in \cite{Embedding of AdS black hole}. Finally, it should be remarked that
the three-form field equation coming from the Lagrangian \eqref{L_7D_from_11D} needs to be supplemented with the odd-dimensional self-duality condition as in the pure $SU(2)$ gauged supergravity discussed in \cite{Pope_N27D_pure}.
\\
\indent The nine scalars, parametrized by
$\tilde{T}_{\alpha\beta}$, in the dimensionally reduced theory are
encoded in the $SL(4,\mathbb{R})/SO(4)$ coset manifold. Therefore,
in order to compare the result with gauged $N=2$ $SO(4)$
supergravity given in the previous section, we need to use the
relation between $SL(4,\mathbb{R})/SO(4)$ and $SO(3,3)/SO(3)\times
SO(3)$ coset manifolds. This is given in \cite{Eric_N2_7Dmassive}.
For the details of this mapping, the reader is referred to
\cite{Eric_N2_7Dmassive}. We will only give the $SO(3,3)/SO(3)\times
SO(3)$ coset representative
$L^A_{\phantom{s}I}=(L^i_{\phantom{s}I},L^r_{\phantom{s}I})$ and
that of $SL(4,\mathbb{R})/SO(4)$, $\mc{V}^\alpha_R$ with
$R=1,\ldots, 4$,
\begin{equation}
L^A_{\phantom{s}I}=\frac{1}{4}\Gamma^{\alpha\beta}_I\eta^A_{RS}\mc{V}^R_{\phantom{s}\alpha}
\mc{V}^S_{\phantom{s}\beta}\label{L_V_relation}
\end{equation}
where $\Gamma^I$ and $\eta^A$ are chirally projected $SO(3,3)$ gamma
matrices.
\\
\indent It can be shown that the scalar potential can be written as
\begin{eqnarray}
V&=&\frac{1}{4}e^{-\sigma}\left(C^{ir}C_{ir}-\frac{1}{9}C^2\right)
+16h^2e^{4\sigma}
-\frac{4\sqrt{2}}{3}he^{\frac{3\sigma}{2}}C\nonumber \\
&=&\frac{1}{8}e^{-\sigma}\left(
T_{\alpha\beta}T_{\alpha\beta}-\frac{1}{2}T_{\alpha\alpha}^2\right)+2T_{\alpha\alpha}he^{\frac{3\sigma}{2}}+16h^2e^{4\sigma}
\end{eqnarray}
This form is similar to the potential \eqref{potential_11D} if
$\tilde{T}_{\alpha\beta}$ is identified with $T_{\alpha\beta}$. Note
that $T_{\alpha\beta}$ and $C$, $C^{ir}$ contain the gauge coupling $g_1$
and $g_2$. In order to compare the Lagrangian of the two theories,
we need to multiply the Lagrangian \eqref{7Daction} by two and
separate the coupling constants $g_1$ and $g_2$ from the structure
constants $f_{IJK}=(g_1\epsilon_{ijk},g_2\epsilon_{rst})$. With
these, the two scalar potentials are exactly the same if we identify
\begin{equation}
g_2=g_1=-16h=-2g\, .
\end{equation}
\indent We also need to redefine the following fields in
the Lagrangian \eqref{7Daction}:
\begin{eqnarray}
& &H_{(4)}\rightarrow \frac{F_{(4)}}{\sqrt{2}},\qquad C_{(3)}\rightarrow \frac{C_{(3)}}{\sqrt{2}},\nonumber \\
& &F^I=\frac{1}{4}\Gamma^I_{\alpha\beta}F^{\alpha\beta}_{(2)}\qquad \textrm{or}\qquad F^{\alpha\beta}_{(2)}=-\frac{1}{2}\epsilon^{\alpha\beta\gamma\delta}\Gamma^I_{\gamma\delta}F^I\nonumber \\
& &X=e^{-\frac{\sigma}{2}}\, .
\end{eqnarray}
By using \eqref{L_V_relation}, it can also be checked that
\begin{equation}
\tilde{T}^{-1}_{\alpha\gamma}\tilde{T}^{-1}_{\beta\delta}=\frac{1}{4}\Gamma^I_{\alpha\beta}\Gamma^J_{\gamma\delta}
\left(L^i_{\phantom{s}I}L_{iJ}+L^r_{\phantom{s}I}L_{rJ}\right).
\end{equation}
The field equations from the two theories also match.
\\
\indent We now move to supersymmetry transformations of fermions.
The maximal $N=4$ theory contains the gravitini $\psi_\mu$ and the
spin-$\frac{1}{2}$ fields $\lambda_{\hat{i}}$. The latter is
decomposed into $(\lambda_R,\lambda_5)$. The $SO(5)_c$
$\Gamma^{\hat{i}}$ gamma matrices are accordingly decomposed as
$\Gamma^{\hat{i}}=(\Gamma^R,\Gamma^5)$.
$\Gamma^5=\Gamma^1\Gamma^2\Gamma^3\Gamma^4$ acts as the chirality
matrix of $SO(4)$. Following \cite{Salam_Sezgin_from10D}, we make
the truncation
\begin{equation}
\epsilon^-=\psi^-_\mu=\lambda_5^-=\lambda^+_\alpha=0\, .
\end{equation}
$\epsilon^\pm$ satisfy $\Gamma^5\epsilon^\pm=\pm\epsilon^\pm$ with
$\epsilon=\epsilon^++\epsilon^-$. We will now drop $\pm$ superscript
from $\epsilon$, $\lambda$ and $\psi_\mu$.
\\
\indent In accordance with the bosonic truncation
$T^{ij}=(T^{\alpha\beta},T^{55})=(X\tilde{T}^{\alpha\beta},X^{-4})$,
we truncate the $SL(5,\mathbb{R})$ coset representative as
$\Pi_i^{\phantom{s}\hat{i}}=(\Pi_\alpha^{\phantom{s}R},\Pi_5^{\phantom{s}\hat{5}})$.
With the identification
$\Pi_\alpha^{\phantom{s}R}=X^{-\frac{1}{2}}\mc{V}_\alpha^{\phantom{s}R}$
and $\Pi_5^{\phantom{s}\hat{5}}=X^{2}$, we can write
$\tilde{T}^{\alpha\beta}$ in term of $SL(4,\mathbb{R})$ coset
representative $\mc{V}_\alpha^{\phantom{s}R}$ as
\begin{equation}
\tilde{T}^{\alpha\beta}=(\mc{V}^{-1})_R^{\phantom{s}\alpha}(\mc{V}^{-1})_S^{\phantom{s}\beta}\delta^{RS},
\qquad \textrm{and}\qquad
\tilde{T}_{RS}=(\mc{V}^{-1})_R^{\phantom{s}\alpha}(\mc{V}^{-1})_S^{\phantom{s}\beta}\delta_{\alpha\beta}\,
.
\end{equation}
\indent We then find that equations \eqref{deltaPsiN4} and
\eqref{deltaLambdaN4} become
\begin{eqnarray}
\delta \psi_\mu &=&D_\mu \epsilon
-\frac{1}{20}g(X\tilde{T}_{RR}+X^{-4})\gamma_\mu\epsilon
-\frac{1}{40\sqrt{2}}X^{-1}\left(\gamma_\mu^{\phantom{s}\nu\rho}-8\delta^\nu_\mu
\gamma^\rho\right)\Gamma_{RS}F^{RS}_{\nu\rho}\epsilon\nonumber \\
&
&-\frac{1}{60}X^{-2}\left(\gamma_\mu^{\phantom{s}\nu\rho\sigma}-\frac{9}{2}
\delta^\nu_\mu\gamma^{\rho\sigma}\right)S^5_{\nu\rho\sigma}\epsilon,\\
\delta \lambda_R &=&\frac{1}{4}\gamma^\mu\Gamma_R X^{-1}\pd_\mu
X\epsilon+\frac{1}{2}\Gamma^S\gamma^\mu P_{RS}\epsilon
+\frac{1}{16\sqrt{2}}X^{-1}\gamma^{\mu\nu}\left(\Gamma_{ST}\Gamma_R-\frac{1}{5}\Gamma_R\Gamma_{ST}\right)
F^{ST}_{\mu\nu}\epsilon\nonumber \\
&
&-\frac{1}{10}gX^{-4}\Gamma_R\epsilon-\frac{1}{2}gX\left(\tilde{T}_{RS}-\frac{1}{5}\tilde{T}_{TT}\delta_{RS}\right)\Gamma^S\epsilon
-\frac{1}{120}X^{-2}\gamma^{\mu\nu\rho}\Gamma_RS^5_{\mu\nu\rho}\epsilon\,
.
\end{eqnarray}
\indent The constraint $\Gamma^{\hat{i}}\lambda_{\hat{i}}=0$ imposes
the condition $\lambda_{5}^+=-\Gamma^R\lambda_R^-$. Therefore, the
independent fields will be $\psi_\mu$ and $\lambda_R$. This is the
reason for excluding $\delta\lambda_{5}$ in the above equations. We
then identify $\Gamma^R\lambda_R$ with $\chi$ and
$\hat{\lambda}_{R}=\lambda_R-\frac{1}{4}\Gamma_R \Gamma^S\lambda_S$
with $\lambda^r$ in \eqref{delta_lambda}. Note that
$\hat{\lambda}_{R}$ has only three independent components due to the
condition $\Gamma^R\hat{\lambda}_R=0$.
\\
\indent With these and the odd-dimensional self-duality, we end up
with, after some gamma matrix algebra,
\begin{eqnarray}
\delta \psi_\mu &=&D_\mu \epsilon -\frac{1}{20}gX\tilde{T}\gamma_\mu
\epsilon
-\frac{1}{40\sqrt{2}}X^{-1}\left(\gamma_\mu^{\phantom{s}\nu\rho}-8\delta^\nu_\mu
\gamma^\rho\right)\Gamma_{RS}F^{RS}_{\nu\rho}\epsilon\nonumber \\
& &-\frac{1}{20}gX^{-4}\gamma_\mu
\epsilon-\frac{1}{480}X^{2}\left(3\gamma_\mu^{\phantom{s}\nu\rho\sigma\tau}-8\delta^\nu_\mu
\gamma^{\rho\sigma\tau}\right)F_{\nu\rho\sigma\tau}\epsilon,\label{delta1}\\
\delta\chi &=&X^{-1}\gamma^\mu\pd_\mu X
\epsilon-\frac{2}{5}gX^{-4}\epsilon+\frac{1}{10}gX\tilde{T}_{RR}\epsilon\nonumber
\\
&
&-\frac{1}{120}X^2\gamma^{\mu\nu\rho\sigma}F_{\mu\nu\rho\sigma}\epsilon-\frac{1}{20\sqrt{2}}X^{-1}
\gamma^{\mu\nu}\Gamma_{RS}F^{RS}_{\mu\nu}\epsilon,\label{delta2}\\
\delta\hat{\lambda}_R&=&-\frac{1}{2}\gamma^\mu\Gamma^S P_{\mu
RS}\epsilon-\frac{1}{8}gX\tilde{T}_{SS}\Gamma_R \epsilon
+\frac{1}{2}gX\tilde{T}_{RS}\Gamma^S\epsilon \nonumber \\
&
&-\frac{1}{8\sqrt{2}}X^{-1}\gamma^{\mu\nu}\Gamma_S\left(F^{RS}_{\mu\nu}
+\frac{1}{2}\epsilon_{RSTU}F^{TU}_{\mu\nu}\right)\epsilon\, .
\label{delta3}
\end{eqnarray}
In the above equations, we have used the following definitions
\begin{eqnarray}
P_{RS}&=&(\mc{V}^{-1})^\alpha_{(R}\left(\delta^\beta_\alpha
d+gA_{(1)\alpha} ^{\phantom{sdss}\beta}\right)\mc{V}_\beta
^{\phantom{s}T}\delta_{S)T},\nonumber \\
Q_{RS}&=&(\mc{V}^{-1})^\alpha_{[R}\left(\delta^\beta_\alpha
d+gA_{(1)\alpha} ^{\phantom{sdss}\beta}\right)\mc{V}_\beta
^{\phantom{s}T}\delta_{S]T},\nonumber \\
D\epsilon &=& d\epsilon
+\frac{1}{4}\omega_{ab}\gamma^{ab}+\frac{1}{4}Q_{RS}\Gamma^{RS}\, .
\end{eqnarray}
\indent Notice that with our convention for
$\Gamma^5\epsilon=\epsilon$, $\Gamma_{RS}$ is anti-self dual. The
field strength $F^{RS}_{(2)}$ appearing in \eqref{delta1} and
\eqref{delta2} must be accordingly anti-self dual. This should be
identified with the $SU(2)$ field strength $F^{i}_{(2)}$ in
\eqref{delta_psi} and \eqref{delta_chi}. On the other hand, the self
dual part of $F^{RS}_{(2)}$ appears in \eqref{delta3} and should be
identified with $F^r_{(2)}$ in \eqref{delta_lambda}.
\\
\indent Using the relation $C=-\frac{3}{2\sqrt{2}}g_1\tilde{T}$ and
identifying $F_{RS}\Gamma^{RS}=-2\sqrt{2}iF^i\sigma^i$, we can see
that equations \eqref{delta1} and \eqref{delta2} match with
equations \eqref{delta_psi} and \eqref{delta_chi} after using the
relation $g_1=-2g$ and gamma matrix identities such as $\gamma_\mu
\gamma^{\nu\rho}=\gamma_\mu^{\phantom{s}\nu\rho}+2\delta_\mu^{[\nu}\gamma^{\rho]}$.
Note that in order to match the gravitino variation, we need to
multiply \eqref{delta1} by two. Comparing \eqref{delta_lambda} and
\eqref{delta3} is more complicated. The $SO(4)$ gamma matrices
$\Gamma^R$ need to be expressed in terms of

\section{Embedding seven-dimensional RG flow to eleven dimensions}\label{uplifting_the_solution}
In this section, we will use the reduction ansatz obtained in the
previous section to uplift some seven-dimensional solutions. The
dimensional reduction gives rise to the condition $g_2=g_1$. This
makes the supersymmetric $AdS_7$ critical point with
$SO(3)_{\textrm{diag}}$ symmetry found in \cite{7D_flow} disappears.
Accordingly, the flow solution given in \cite{7D_flow} cannot be
uplifted to eleven dimensions with the present reduction ansatz.
However, to give examples of the uplifted solutions, we will study
other solutions in the case of $g_2=g_1$.
\subsection{Uplifting $AdS_7$ solutions}
We now further truncate the nine scalars given by
$\tilde{T}_{\alpha\beta}$ to one scalar invariant under
$SO(3)_{\textrm{diag}}\subset SO(3)\times SO(3)\sim SO(4)$. This
scalar sector has already been studied in \cite{7D_flow}. We will
give more solutions in this section. Under $SO(3)_{\textrm{diag}}$,
the nine scalars transform as $\mathbf{1}+\mathbf{3}+\mathbf{5}$.
There is only one singlet. It can be checked that the
$SO(3)_{\textrm{diag}}$ singlet correspond to
\begin{equation}
\mc{V}^R_{\phantom{s}\alpha}=\left(\begin{array}{cccc}
                       e^{\frac{\phi}{2}} &  &  &  \\
                        & e^{\frac{\phi}{2}} &  &  \\
                        &  & e^{\frac{\phi}{2}} &  \\
                        &  &  & e^{-\frac{3\phi}{2}} \\
                     \end{array}
                   \right)\qquad
                   \textrm{or}
                   \qquad
\tilde{T}_{\alpha\beta}=\left(
                     \begin{array}{cccc}
                       e^{\phi} &  &  &  \\
                        & e^{\phi} &  &  \\
                        &  & e^{\phi} &  \\
                        &  &  & e^{-3\phi} \\
                     \end{array}
                   \right).
\end{equation}
$\tilde{T}_{\alpha\beta}$ can be written more compactly as
$\tilde{T}_{\alpha\beta}=(\delta_{ab}e^{\phi},e^{-3\phi})$ for
$a,b=1,2,3$. By using \eqref{L_V_relation} and the explicit form of
$\Gamma^I$ and $\eta^A$ given in \cite{Eric_N2_7Dmassive}, it is
easy to verify that this $\mc{V}$ precisely gives the
$SO(3,3)/SO(3)\times SO(3)$ coset representative $L$ used in
\cite{7D_flow}.
\\
\indent Using this and the relation $X=e^{-\frac{\sigma}{2}}$, we
find the scalar potential
\begin{equation}
V=\frac{1}{2}g^2e^{-\sigma}\left[e^{5\sigma+e^{-6\phi}}-6e^{-2\phi}-3e^{2\phi}-2e^{\frac{5}{2}\sigma-3\phi}
\left(1+3e^{4\phi}\right)\right].
\end{equation}
This potential admits two $AdS_7$ critical points given by
\begin{eqnarray}
\sigma &=&\phi=0,\qquad V_0=-480h^2\\
\sigma &=&-\frac{1}{10}\ln 2,\qquad \phi=-\frac{1}{4}\ln 2,\qquad
V_0=-160\times 2^{\frac{3}{5}}h^2
\end{eqnarray}
where we have used $g=8h$ or equivalently $g_1=-16h$ as given in
\cite{7D_flow}. By using the BPS equations given in \cite{7D_flow},
which are repeated below, we see that the second critical point is
non-supersymmetric. Scalar masses at this critical point can be
computed to be
\\
\begin{center}
\begin{tabular}{|c|c|}
  \hline
  $SO(3)_{\textrm{diag}}$ & $m^2L^2$ \\ \hline
  $\mathbf{1}$ & $-12$ \\
  $\mathbf{1}$ & $12$ \\
  $\mathbf{3}$ & $0$ \\
  $\mathbf{5}$ & $-12$ \\
  \hline
\end{tabular}
\end{center}
where the $AdS_7$ radius is given by $L=\sqrt{-\frac{15}{V_0}}$. The
three massless scalars are the expected Goldstone bosons
corresponding to the symmetry breaking of $SO(4)$ to $SO(3)$. One of
the $\mathbf{1}$ and $\mathbf{5}$ scalars have masses below the BF
bound $m^2L^2=-9$, so this critical point is unstable.
\\
\indent The first critical point is the trivial point preserving all
supersymmetries and the full $SO(4)$ gauge symmetry. The scalar
masses can be found in \cite{7D_flow}. We will now uplift this
$AdS_7$ vacuum to eleven dimensions. We begin with the coordinates
$\mu^\alpha =(\cos\psi\hat{\mu}^a,\sin\psi)$ in which
$\hat{\mu}^a\hat{\mu}^a=1$. Since $\sigma=\phi=0$, we then find
$\Delta=1$ and
\begin{eqnarray}
ds^2_{11}&=&e^{\frac{2r}{L_{UV}}}dx^2_{1,5}+dr^2+\frac{1}{32h^2}\left[d\xi^2+\frac{1}{4}\cos^2\xi \left(d\psi^2+\cos^2\psi d\Omega^2_2\right)\right]\label{AdS7_in_11D}\\
\hat{F}_{(4)}&=&-\frac{3}{256h^3}\cos^5\xi d\xi \wedge
\epsilon_{(3)}
\end{eqnarray}
where $d\Omega_2^2$ is the metric on the two-sphere. The eleven
dimensional geometry is given by $AdS_7\times S^4$. Turning on the
dilaton $\sigma$ would deform the four-sphere but leave the $S^3$
inside invariant. If $\phi,\sigma\neq 0$, the metric would be
further deformed in such a way that the $S^2$ part described by
$d\Omega^2_2$ is invariant. The unbroken symmetry in this case is
the $SO(3)$ isometry of this $S^2$ identified with the unbroken
$SO(3)_\textrm{diag}$. The $SO(3)$ critical point is however
unstable. Therefore, we will not consider $AdS_7$ solution with
$SO(3)$ symmetry.

\subsection{Uplifting RG flows to non-conformal $SO(3)$ Super Yang-Mills}
To give more examples, we will study RG flow solutions to
non-conformal Super Yang-Mills theories in the IR. We will work in
the theory of section \ref{7D_gaugedN2}. With $g_2=g_1$ and the
standard domain wall metric ansatz $ds^2_7=e^{A(r)}dx^2_{1,5}+dr^2$,
the BPS equations taken from \cite{7D_flow} become
\begin{eqnarray}
\phi'&=&-4e^{-\frac{\sigma}{2}-3\phi}\left(e^{4\phi}-1\right)h,\\
\sigma'&=&\frac{8}{5}e^{-\frac{\sigma}{2}-3\phi}\left(1+3e^{4\phi}-4e^{\frac{5}{2}\sigma+3\phi}\right)h,\\
A'&=&\frac{4}{5}e^{-\frac{\sigma}{2}-3\phi}\left(1+3e^{4\phi}+e^{\frac{5}{2}\sigma+3\phi}\right)
\end{eqnarray}
in which $\frac{d}{dr}$ is denoted by $'$. After changing to the new
coordinate $\tilde{r}$ given by
$\frac{d\tilde{r}}{dr}=e^{-\frac{\sigma}{2}}$, we find the solution
\begin{eqnarray}
16h\tilde{r}&=&\ln \left[\frac{1+e^\phi}{1-e^\phi}\right]-2\tan^{-1}\phi +C_1,\\
\sigma &=&\frac{2}{5}\left[\phi-\ln \left[1+12C_2-12C_2e^{4\phi}\right]\right],\\
A&=&\frac{1}{4}\left[\phi-2\ln
(1-e^{4\phi})\right]-\frac{1}{8}\sigma \, .
\end{eqnarray}
The solution interpolates between an $AdS_7$ in the UV,
$\tilde{r}\sim r\rightarrow \infty$, and a domain wall in the IR,
$4h\tilde{r}\rightarrow \tilde{C}$, for a constant $\tilde{C}$.
\\
\indent At the UV, the solution becomes
\begin{equation}
\sigma\sim \phi \sim e^{-16hr}\sim e^{-\frac{4r}{L_{UV}}},\qquad
A\sim 4hr\sim \frac{r}{L_{UV}}\, .
\end{equation}
The eleven-dimensional metric is given by \eqref{AdS7_in_11D}.
\\
\indent In the IR, we find that $\phi$ blows up as
\begin{equation}
\phi\sim -\ln (4h\tilde{r}-\tilde{C})
\end{equation}
for a constant $\tilde{C}$. The behaviour of $\sigma$ depends on the
value of the integration constant $C_2$.
\\
\indent For $C_2=0$, we find
\begin{equation}
\sigma\sim -\frac{2}{5}\ln (4h\tilde{r}-\tilde{C})\sim
-\frac{1}{2}\ln(4hr-C)
\end{equation}
where we have used the relation between $\tilde{r}$ and $r$ in the
IR limit with $C$ being another integration constant. The
seven-dimensional metric is given by
\begin{equation}
ds^2_7=(4hr-C)^2dx^2_{1,5}+dr^2\, .
\end{equation}
\indent For $C_2\neq 0$, the solution becomes
\begin{eqnarray}
\sigma &\sim & \frac{6}{5}\ln (4h\tilde{r}-\tilde{C})\sim \frac{3}{4}\ln (4hr-C),\nonumber \\
ds^2_7&=&(4hr-C)^\frac{3}{4}dx^2_{1,5}+dr^2\, .
\end{eqnarray}
Both cases give $V\rightarrow -\infty$, so the solution is physical
by the criterion of \cite{Gubser_singularity}.
\\
\indent We now look at the eleven-dimensional geometry. For $C_2=0$
and $C_2\neq 0$, the eleven dimensional metric is given respectively
by
\begin{eqnarray}
ds^2_{11}&=&\left(1-\sin^2 \xi \cos^2
\psi\right)^{-\frac{1}{3}}\left[\left(\frac{14}{3}h\rho\right)^2dx^2_{1,5}+d\rho^2\right]
+\frac{1}{32h^2}\left(1-\sin^2\xi
\cos^2\psi\right)^{-\frac{2}{3}}\times\nonumber
\\
& &\left[\left(\frac{14}{3}h\rho\right)^{-\frac{27}{7}}\sin^2\xi
\cos^2\psi d\xi^2 +\frac{1}{4}\sin\xi
\sin(2\psi)\left(\frac{14}{3}h\rho\right)^{-\frac{1}{2}}d\psi
d\xi\right. \nonumber \\
&
&\left.+\frac{1}{4}\left(\frac{14}{3}h\rho\right)^{-\frac{20}{7}}d\psi^2+\frac{1}{4}\cos^2\psi
\left(\frac{14}{3}h\rho\right)^{\frac{10}{7}}d\Omega_2^2 \right],\\
ds^2_{11}&=&(\cos\xi
\cos\psi)^{-\frac{2}{3}}\left[\left(\frac{14}{3}h\rho\right)^{\frac{13}{14}}dx^2_{1,5}+d\rho^2\right]
+\frac{1}{32h^2}(\cos\xi \cos \psi)^{-\frac{4}{3}}\times\nonumber
\\
&
&\left[\left(\frac{14}{3}h\rho\right)^{\frac{17}{14}}\left(1-\sin^2\xi
\cos^2\psi\right)d\xi^2 -\frac{1}{4}\sin\xi
\sin(2\psi)\left(\frac{14}{3}h\rho\right)^{\frac{7}{4}}d\xi
d\psi\right.\nonumber \\
& &\left.+\frac{1}{4}\cos^2\xi
\left(\frac{14}{3}h\rho\right)^{\frac{10}{7}}\left(\sin^2\psi
d\psi^2+\cos^2\psi d\Omega_2^2\right)\right]
\end{eqnarray}
where $\left(\frac{14}{3}h\rho\right)^{\frac{6}{7}}=4hr-C$.
\\
\indent As expected, when turning on $\phi$ and $\sigma$, the warped
factors involve coordinates $(\xi,\psi)$. The $S^4$ is then deformed
leaving the $S^2$ intact. If only $\sigma\neq 0$, the $S^3$ part of
the internal metric would be invariant as pointed in
\cite{Pope_N27D_pure}. The deformation with only $\phi\neq 0$ is not
possible since the BPS equation for $\sigma$ would imply $\phi=0$ as
pointed out in \cite{7D_flow}.
\section{Conclusions}\label{conclusion}
In this paper, we have constructed $N=2$ $SO(4)$ gauged supergravity
in seven dimensions with topological mass term. The resulting theory
admit $AdS_7$ vacua and could be useful in the context of the
AdS/CFT correspondence. The resulting reduction ansatz has been
found by truncating the $S^4$ reduction leading to $N=4$ $SO(5)$
gauged supergravity and can be used to uplift seven-dimensional
solutions to eleven dimensions. We have also constructed new
seven-dimensional RG flow solutions and uplifted the resulting
solutions to eleven dimensions. The flows can be interpreted as
deformations of the UV $N=(1,0)$ SCFT in six dimensions with $SO(4)$
symmetry to non-conformal SYM with $SO(3)_{\textrm{diag}}$ symmetry.
These deformations are driven by vacuum expectation values of
dimension 4 operators. Additionally, the result of this paper can be
used to uplift flows to $SO(2)$ non-conformal gauge theories studied
in \cite{7D_flow} for $g_2=g_1$.
\\
\indent However, the RG flow between two supersymmetric $AdS_7$
critical points recently found in \cite{7D_flow} cannot be uplifted
by using the reduction ansatz constructed here. It would be
interesting to find an embedding of this solution in 10 or 11
dimensions. It is also interesting to extend the reduction ansatz
given here to non-compact gauge groups $SO(3,1)$ and $SO(2,2)$. The
internal manifold should involve hyperbolic spaces $H^{3,1}$ and
$H^{2,2}$, respectively. Other possible non-compact gauge groups are
$SL(3,\mathbb{R})$, $SO(2,1)$ and $SO(2,2)\times SO(2,1)$. It would
be very interesting to find higher dimensional origins for these
gauge groups as well. Finally, more insight to six-dimensional gauge
theories might be gained from studying these seven-dimensional
gauged supergravities via AdS$_7$/CFT$_6$ correspondence. We hope to
come back to these issues in future works.

\acknowledgments The author gratefully thanks Eric Bergshoeff for
useful correspondences. This work is supported by Chulalongkorn
University through Ratchadapisek Sompote Endowment Fund under grant
Sci-Super 2014-001. The author is also supported by The Thailand
Research Fund (TRF) under grant TRG5680010.



\begin{thebibliography}{99}
\bibitem{Pope_sphere} M. Cvetic, H. Lu and C. N. Pope, ``Consistent Kaluza-Klein sphere
reductions'', Phys. Rev. \textbf{D62} (2000) 064028, arXiv:
hep-th/0003286.
\bibitem{deWit_S7} B. de Wit and H. Nicolai,
``The consistency of the $S^7$ truncation in $D=11$ supergravity'',
Nucl. Phys. \textbf{B281} (1987) 211.
\bibitem{S4_reduction_11D} H. Natase, D. Vaman and P. van Nieuwenhuizen, ``Consistency of the $AdS_7\times S_4$
reduction and the origin of self-duality in odd dimensions'', Nucl.
Phys. \textbf{B581} (2000) 179-239, arXiv: hep-th/9911238.
\bibitem{Pope_S5_reduction} M. Cvetic, H. Lu, C. N. Pope, A. Sadrzadeh and T. A. Tran,
``Consistent $SO(6)$ reduction of type IIB supergravity on $S^5$'',
Nucl. Phys. \textbf{B586} (2000) 275-286, arXiv: hep-th/0003103.
\bibitem{maldacena} J. M. Maldacena, ``The large $N$ limit of
superconformal field theories and supergravity'', Adv. Theor. Math.
Phys. \textbf{2} (1998) 231-252, arXiv: hep-th/9711200.
\bibitem{LargeN_2_0} R. G. Leigh and M. Rozali, ``The large N limit of the $(2,0)$ superconformal field theory'', Phys. Letts \textbf{B431} (1998) 311-316, arXiv: hep-th/9803068.
\bibitem{Berkooz_6D_dual} M. Berkooz, ``A supergravity dual of a $(1,0)$ field theory in six
dimensions'', Phys. Lett. \textbf{B437} (1998) 315-317, arXiv:
hep-th/9802195.
\bibitem{AdS7_orbifold1} C. Ahn, K. Oh and R. Tatar, ``Orbifolds $AdS_7\times S^4$ and six-dimensional $(0,1)$
SCFT'', Phys. Lett. \textbf{B442} (1998) 109-116, arXiv:
hep-th/9804093.
\bibitem{AdS7_orbifold2} E. G. Gimon and C. Popescu, ``The operator spectrum of the six-dimensional $(1,0)$
theory'', JHEP 04 (1999) \textbf{018}, arXiv: hep-th/9901048.
\bibitem{Ferrara_AdS7CFT6} S. Ferrara, A. Kehagias, H. Partouche and
A. Zaffaroni, ``Membranes and fivebranes with lower supersymmetry
and their AdS supergravity duals'', Phys. Lett. \textbf{B431} (1998)
42-48, arXiv: hep-th/9803109.
\bibitem{Pure_N2_7D1} P. K. Townsend and P. van Nieuwenhuizen, ``Gauged seven-dimensional
supergravity'', Phys. Lett. \textbf{B125} (1983) 41-46.
\bibitem{Eric_N2_7D} E. Bergshoeff, I. G. Koh and E. Sezgin, ``Yang-Mills-Einstein supergravity in seven
dimensions'', Phys. Rev. \textbf{D32} (1985) 1353-1357.
\bibitem{Park_7D} Y. J. Park, ``Gauged Yang-Mills-Einstein supergravity with three index field in seven
dimensions'', Phys. Rev. \textbf{D38} (1988) 1087.
\bibitem{Salam_7DN2} A. Salam and E. Sezgin, ``$SO(4)$ gauging of $N=2$ supergravity in
seven-dimensions'', Phys. Lett. \textbf{B126} (1983) 295.
\bibitem{Eric_N2_7Dmassive} E. Bergshoeff, D. C. Jong and E. Sezgin, ``Noncompact gaugings,
chiral reduction and dual sigma model in supergravity'', Class.
Quant. Grav. \textbf{23} (2006) 2803-2832, arXiv: hep-th/0509203.
\bibitem{7D_flow} P. Karndumri, ``RG flows in 6D $N=(1,0)$ SCFT from $SO(4)$ half-maximal gauged supergravity''
, JHEP 06 (2014) \textbf{101}, arXiv: 1404.0183.
\bibitem{Pope_N27D_pure} H. Lu and C. N. Pope ``Exact embedding of $N=1$, $D=7$ gauged supergravity in $D=11$'',
Phys. Letts. \textbf{B467} (1999) 67-72, arXiv: hep-th/9906168.
\bibitem{Salam_Sezgin_from10D} M. Cvetic, G. W. Gibbons and C. N. Pope,
``A String and M-theory Origin for the Salam-Sezgin Model'', Nucl.
Phys. \textbf{B667} (2004) 164-180, arXiv: hep-th/0308026.
\bibitem{S3_S4_reduction_IIA} M. Cvetic, H. Lu, C. N. Pope, A. Sadrzadeh and T. A. Tran,
``$S^3$ and $S^4$ reductions of type IIA supergravity'', Nucl. Phys.
\textbf{B590} (2000) 233-251, arXiv: hep-th/0005137.
\bibitem{Peter_S4} H. Nastase, D. Vaman and P. van Nieuwenhuizen,
``Consistent nonlinear KK reduction of 11d supergravity on
$AdS_7\times S_4$ and self-duality in odd dimensions'', Phys. Lett.
\textbf{B469} (1999) 96-102, arXiv: hep-th/9905075.
\bibitem{Natase_S4} H. Nastase and D. Vaman, ``On the nonlinear KK reductions on spheres of supergravity
theories'', Nucl. Phys. \textbf{B583} (2000) 211-236, arXiv:
hep-th/0002028.
\bibitem{7DN4_Peter} M. Pernici, K. Pilch and P. van Nieuwenhuizen,
``Gauged maximally extended supergravity in seven dimensions'',
Phys. Lett. \textbf{B143} (1984) 103.
\bibitem{Embedding of AdS black hole} M. Cvetic, M. J. Duff, P. Hoxha, James T. Liu,
H. Lu, J. X. Lu, R. Martinez-Acosta, C. N. Pope, H. Sati and T. A. Tran, ``Embedding AdS Black Holes in
Ten and Eleven Dimensions'', Nucl. Phys. \textbf{B558} (1999) 96-126, arXiv: hep-th/9903214.
\bibitem{Gubser_singularity} S. S. Gubser, ``Curvature singularities: the good, the bad and the naked'', Adv. Theor.
Math. Phys. \textbf{4} (2000) 679-745.
\end{thebibliography}
\end{document}